\documentclass[reprint, aip, apl, amsmath,amssymb,superscriptaddress]{revtex4-1}

\usepackage{graphicx}
\usepackage{dcolumn}
\usepackage{bm}
\usepackage[sort&compress]{natbib}

\begin{document}

\title{Independent electrical tuning of separated quantum dots in coupled photonic crystal cavities}

\author{S. M. Thon}
\altaffiliation{Current Address: Department of Electrical and Computer Engineering, University of Toronto, 10 King's College Road, Toronto, Ontario M5S 3G4, Canada}
\affiliation{Department of Physics, University of California Santa Barbara, Santa Barbara, California 93106, USA}
\author{H. Kim}
\altaffiliation{Current Address: Department of ECE, IREAP, University of Maryland, College Park, Maryland 20742, USA}
\affiliation{Department of Physics, University of California Santa Barbara, Santa Barbara, California 93106, USA}
\author{C. Bonato} \email{bonato@physics.leidenuniv.nl}
\affiliation{Huygens Laboratory, Leiden University, P.O. Box 9504, 2300 RA Leiden, the Netherlands}
\author{J. Gudat}
\affiliation{Huygens Laboratory, Leiden University, P.O. Box 9504, 2300 RA Leiden, the Netherlands}
\author{J. Hagemeier}
\affiliation{Department of Physics, University of California Santa Barbara, Santa Barbara, California 93106, USA}
\author{P. M. Petroff}
\affiliation{Materials Department, University of California Santa Barbara, Santa Barbara, California 93106, USA}
\affiliation{ECE Department, University of California Santa Barbara, Santa Barbara, California 93106, USA}
\author{D. Bouwmeester}
\affiliation{Department of Physics, University of California Santa Barbara, Santa Barbara, California 93106, USA}
\affiliation{Huygens Laboratory, Leiden University, P.O. Box 9504, 2300 RA Leiden, the Netherlands}

\date{\today}

\begin{abstract}
Systems of photonic crystal cavities coupled to quantum dots are a promising architecture for quantum networking and quantum simulators.  The ability to independently tune the frequencies of laterally separated quantum dots is a crucial component of such a scheme.  Here, we demonstrate independent tuning of laterally separated quantum dots in photonic crystal cavities coupled by in-plane waveguides by implanting lines of protons which serve to electrically isolate different sections of a diode structure.
\end{abstract}

\maketitle

Quantum dots coupled to optical microcavities represent a viable candidate for integrated quantum information technologies such as single photon sources and quantum memory/repeater networks\cite{ChildressPRL06,CiracPRL97, bonatoPRL10}. A long-term goal would be to combine on a single chip multiple cavities, each one embedding a single emitter, and connect them through waveguides. Such a system would be a basic building block for a scalable quantum information processing architecture, allowing the implementation of multi-atom entangled states via photon manipulation \cite{choPRL05, choPRA08, rarityPRB08}. Furthermore, coupled cavity arrays have been proposed as a quantum simulation tool to investigate the dynamics of quantum many-body systems, originally encountered in condensed-matter physics (like the Bose-Hubbard or Heisenberg models) \cite{hartmannNP06, hartmannLPR08}.

Photonic crystals are ideal platforms for such a system because defect cavities can easily be integrated with waveguides in the same planar photonic lattice structures \cite{FaraonOE08,FaraonAPL07b}. Coupling between the cavities and waveguides can be turned off and on relatively easily by, for instance, modifying the local refractive index in a waveguide region via a localized intense laser pulse\cite{FushmanAPL07}. \\
A significant barrier to practical implementation of coupled cavity-quantum dots systems is, however, the possibility to independently tune the quantum dot wavelengths in different cavity regions.  Because of the large quantum dot ensemble frequency spread of self-assembled InGaAs quantum dots during the growth process, independent frequency control of spatially separated quantum dots is a vital tool.  Currently existing methods designed to independently tune different quantum dots have significant limitations. Temperature tuning via local heating \cite{FaraonAPL07} requires large distances between the QDs for thermal separation and tends to degrade the quantum dot optical quality at high temperature.  Lateral electric field tuning of quantum dots in Schottky diode devices has also been studied \cite{faraonNJP11}, but lateral electric fields provide limited QD emission frequency tuning ranges and do not allow for control of the QD charging states. An ideal tuning mechanism would be all-electrical in implementation and have the flexibility to address quantum dots on any area of the chip with only a small modification of the standard fabrication process.

Here, we demonstrate such a mechanism, by implanting protons in a localized area which electrically isolates regions of quantum dots embedded in a diode structure. Ion or proton implantation is used in semiconductor processing as an isolating technique, since it causes free carrier compensation in a doped semiconductor layer through either irradiation-induced damage or chemically-related deep levels \cite{ridgwayNIM93}.  We show that quantum dots in two spatially-separated  waveguide-coupled cavities can be tuned independently, and we also demonstrate that the cavities are optically coupled by pumping one and detecting emission in the other.  Our tuning mechanism is robust and scalable, and could be used to independently tune quantum dots laterally separated by as little as 3 $\mu m$.

\begin{figure}[h]
    \includegraphics[width=8.5cm, clip=true]{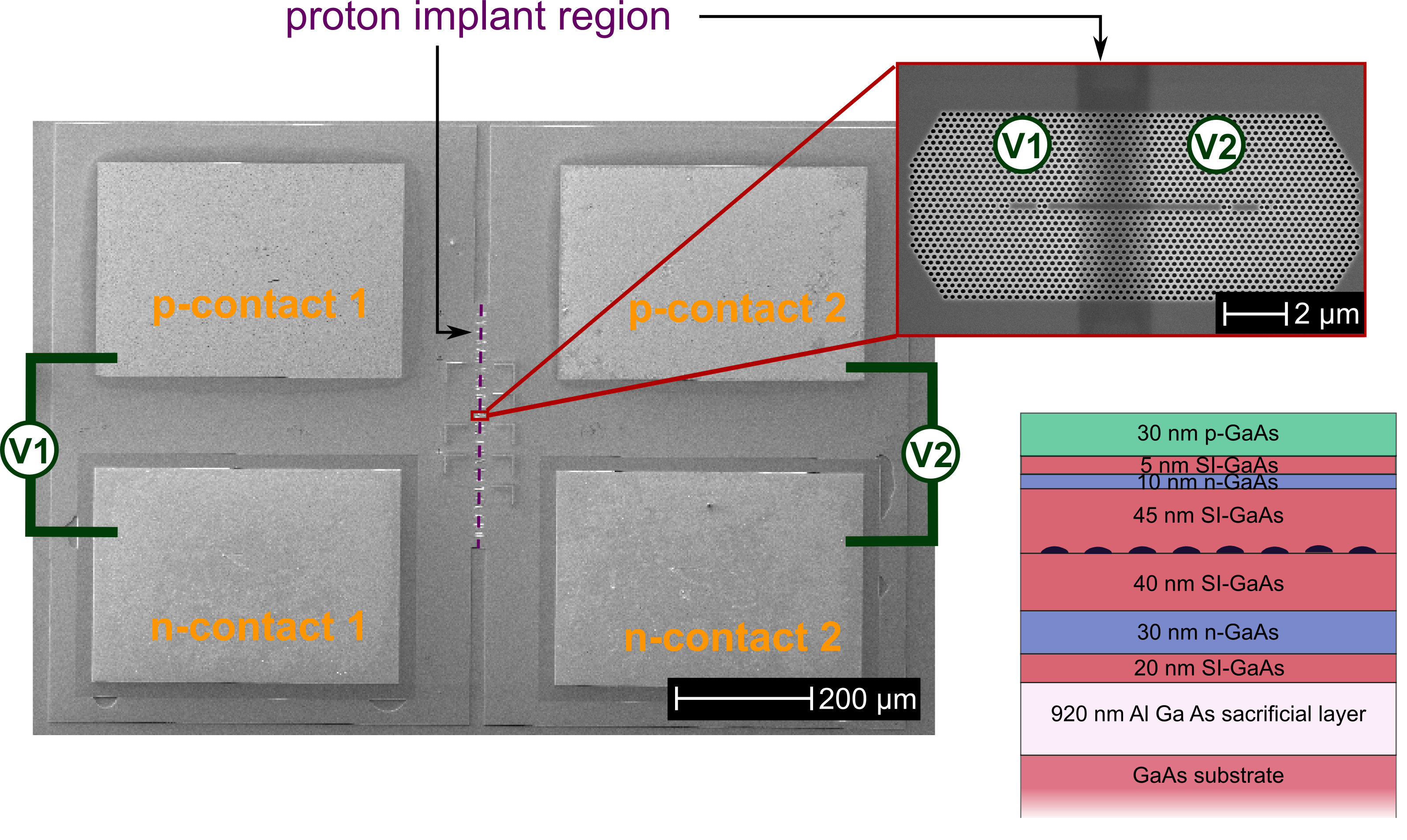}
    \caption{Left: SEM image of the two n- and p-contact pads with the proton implant region and photonic crystal devices in between (labeled).  Top right: SEM image of one of the photonic crystal devices.  The proton implant region is clearly visible as a dark strip through the center of the device.  Bottom right: schematic of the layer structure for the proton implant sample.}
    \label{fig:pilayout}
\end{figure}

Our sample consists of a 180 nm GaAs membrane grown by molecular beam epitaxy on top of a 0.92 $\mu$m Al$_{0.7}$Ga$_{0.3}$As sacrificial layer on a GaAs substrate. The In$_{0.4}$Ga$_{0.6}$As quantum dot layer is grown in the center of the GaAs membrane by depositing 10 periods of 0.55 {\AA} of InAs and 1.2 {\AA} of In$_{0.13}$Ga$_{0.87}$As. We used a p-i-n diode design by growing doped layers within the membrane: 30~nm thick p-doped ($2\times 10^{19}/\textrm{cm}^3$) and n-doped ($2\times 10^{18}/\textrm{cm}^3$) GaAs layers are grown as the top and bottom layers of the GaAs membrane, respectively.  As shown in Fig.~\ref{fig:pilayout}, a thin n-doped layer is introduced below the top p-doped layer, which reduces the electric field across the quantum dot by flattening the band structure. In this modified structure, quantum dot emission is visible al lower voltages than in a simple p-i-n diode structure.

We fabricated two sets of electrical contacts in order to test the proton implant isolation barrier and independent tuning scheme.  Because the doped layers are very thin, the sample must be etched to nanometer accuracy to make good electrical contacts to the buried n-doped layer.  We use a citric acid/H$_2$O$_2$ etch to expose the n-doped layer.  The n-contact metal consists of Ni/AuGe/Ni/Au which is evaporated and annealed.  The p-contact metal is evaporated Ti/Pt/Au.  After the contacts are defined, we deposit 200~nm of Si$_3$N$_4$ on the sample to protect the surface during the proton implant procedure and to maximize the number of protons delivered to the p-doped layer at the specific calculated implant energy.   We then perform a final photolithography step to expose the areas in which the proton implant isolating barriers will be created. We defined strips of 1 and 2 $\mu$m width in between the two sets of electrical contacts, while the rest of the sample area remained protected by photoresist.  To effectively remove the carriers in the p-doped layer, the proton implant dose and energy are calculated using the TRIM (Transport of Ions in Matter) simulator\cite{Ziegler85}.  We performed a proton implantation using a dose of $5\times 10^{14}~\textrm{cm}^{-2}$ and implant energy of 35 keV in order to electrically isolate the two sets of contacts.

After proton implantation, we fabricated columns of photonic crystal devices across the proton implantation region, using a standard electron beam lithography process, followed by reactive ion etching to define the pattern in the membrane and a HF undercut etch to remove the sacrificial layer.  The sample layout is shown in Fig.~\ref{fig:pilayout}.  Our photonic crystal devices consisted of two L3-type defect cavities separated by approximately 7.5 $\mu$m with a waveguide in between, similar to the devices studied by A. Faraon et al. \cite{FaraonAPL07b}.  The proton implant region cuts through the approximate center of the waveguide in order to electrically isolate the two cavities.

\begin{figure}[h]
    \includegraphics[width=8.5cm, clip=true]{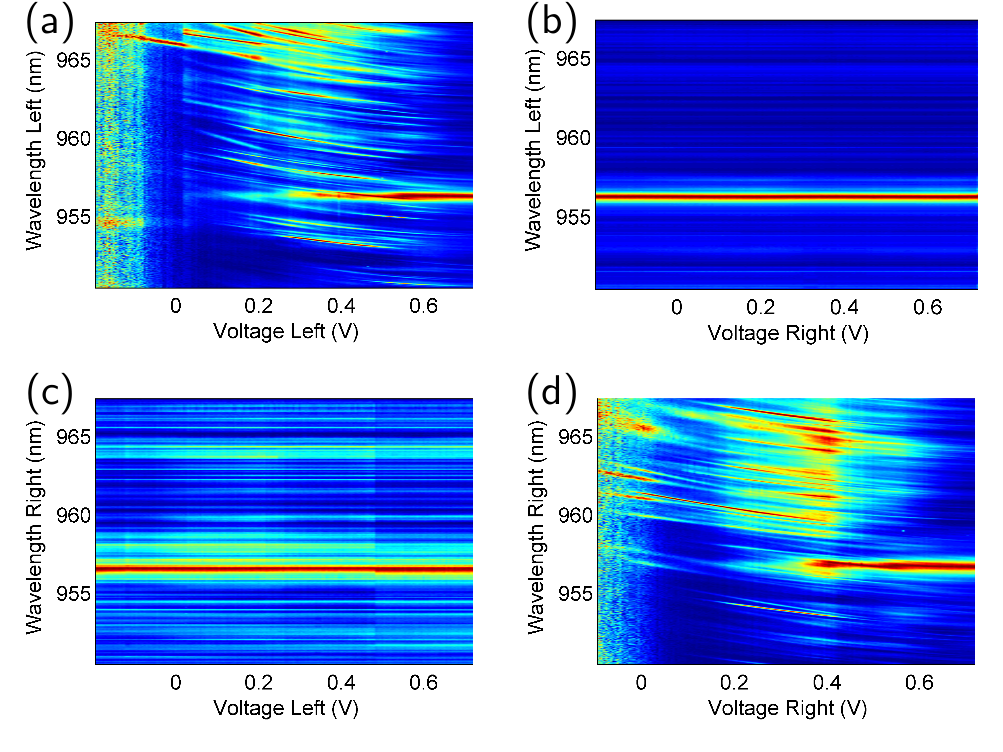}
    \caption{Each of the two cavity regions is monitored separately as the voltage is scanned on either the emission cavity or the opposite cavity. (a) The voltage is scanned across the left set of contacts while the emission from the left cavity is monitored.  Clear tuning of quantum dot lines is visible. (b) The voltage is scanned across the right set of contacts while the emission from the left cavity is monitored.  No tuning is visible.  (c) The voltage is scanned across the left set of contacts while the emission from the right cavity is monitored.  No tuning is visible.  (d) The voltage is scanned across the right set of contacts while the emission from the right cavity is monitored.  Clear tuning of quantum dot lines is visible.  Each spectrum in (a)-(d) is normalized to itself so that the maximum intensity remains constant across each scan.}
    \label{fig:pivscan}
\end{figure}

The sample was mounted in a He-flow cryostat and pumped by a laser at 830 nm in a standard photoluminescence setup.  The emission was collected through the same microscope objective and coupled into a single mode fiber for spatial selectivity.  The output of the single mode fiber was sent through a spectrometer and detected on a CCD.  We tested the tuning mechanism by keeping one side of the device at a constant voltage and scanning the voltage across the other side of the device.  We pumped and monitored the emission from each of the two cavities, taking four scans in total.  The results are shown in Fig.~\ref{fig:pivscan}.  There were many dots in the cavity region for this particular device.  When the voltage was scanned across the detected cavity, clear Stark shift tuning of the quantum dot emission lines was observed over 2 nm range; when the voltage was scanned across the opposite side of the device, the spectrum showed no voltage dependence.  The same was true for all measured devices, proving that the proton implantation served to completely electrically isolate the two contact regions.

In order to probe the coupling between the two waveguide-separated cavities, we performed a series of spatial scans on the sample.  We modified our photoluminescence setup by mounting a dichroic mirror which reflected the pump wavelength and transmitted the cavity mode wavelength immediately in front of the input to the microscope objective.  The dichroic mirror's tip and tilt axes were controlled by motorized actuators.  This allowed us to controllably tilt the pump beam relative to the input of the microscope objective, thereby spatially separating the pump and collection locations on the sample.  We kept the collection location constant on one of the cavities and scanned the pump beam around the entire 2D plane of the photonic crystal device while taking photoluminescence spectra.  In this way, we were able to map out the cavity emission as a function of pump position.

\begin{figure}[h]
    \includegraphics[width=8.5cm, clip=true]{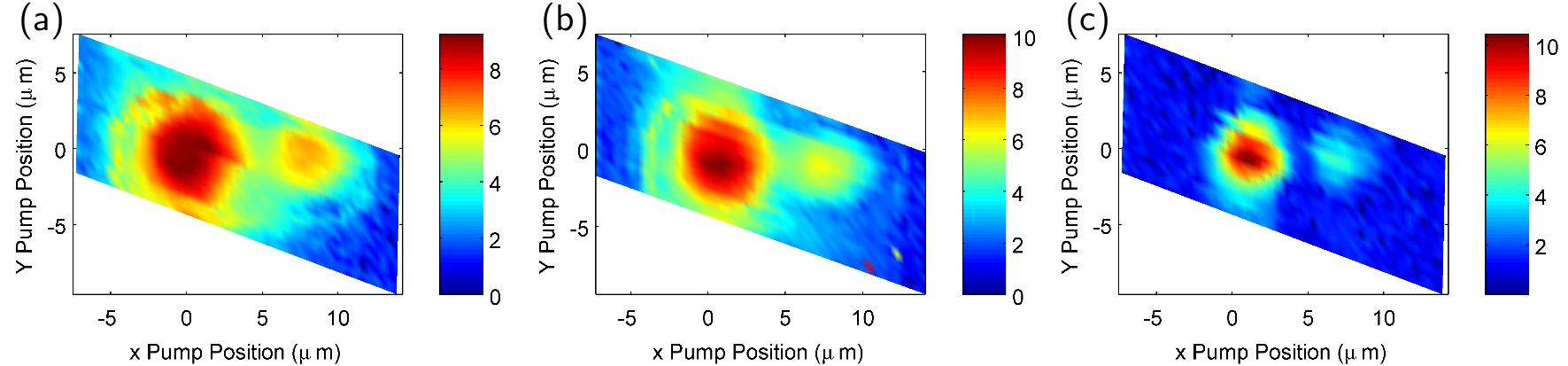}
    \caption{(a)-(c) Spatial maps of the emission from the left cavity of three different devices at the left cavity resonant wavelength as a function of pump position on the sample.  The color scale corresponds to log of intensity.  (a) and (b) are scans of cavities in which the cavity mode resonant wavelengths are spaced by less than their linewidths.  (c) is a scan of a cavity with the two cavity mode wavelengths spaced by more than their linewidths.}
    \label{fig:picavscan}
\end{figure}

The results for three different devices are shown in Fig. ~\ref{fig:picavscan}, with the color map scale set by the log of the intensity at the resonant frequency of the left cavity mode.  The three scans display similar characteristics.  Each has a bright peak on the left side which corresponds to pumping and collecting light from the left cavity.  Each also has a secondary, weaker peak on the right side which corresponds to pumping the right cavity and collecting light from the left cavity.  However, the relative intensities of the two peaks are different in the three scans.  The scans in \ref{fig:picavscan}(a) and (b) are of devices in which the two cavity mode frequencies are spaced by less than the linewidths of the cavity modes.  The scan in \ref{fig:picavscan}(c) is of a device in which the cavity mode frequencies are spaced by much more than the linewidths of the cavity modes.  In \ref{fig:picavscan}(a), the peak intensity in the secondary bright spot is 7\% of the peak intensity in the primary bright spot.  In \ref{fig:picavscan}(b), the peak intensity ratio is 2\% whereas in \ref{fig:picavscan}(c) the peak intensity ratio is only 0.2\%.

The coupling between the cavities could be optimized in future designs. For example, the geometry of the structure could be altered by changing the cavity-waveguide angle or optimizing the length of the waveguide.  In addition, reducing the size of the holes immediately next to the waveguide could improve the coupling efficiency by ensuring that the coupling occurs in the dispersion-free linear region of the waveguide band \cite{FaraonAPL07}.\\
In order to have sufficiently strong coupling between each emitter and its cavity, the emitter must be precisely located in a spatial maximum of a cavity mode field intensity. This could be done by positioning each cavity around a randomly nucleated quantum dot \cite{ThonAPL09}. Alternatively, one could grow regular arrays of site-controlled quantum dots\cite{schneiderAPL08}, by defining nanoholes as nucleation centers, and align the coupled cavities to waveguides on the array.

In conclusion, we have demonstrated a technique for independently tuning laterally separated quantum dots.  Our method can be integrated into current fabrication processes with only two additional fabrication steps and can be used to independently tune dots that are laterally separated by as little as 2--3 $\mu$m.  In addition, we have used this method to demonstrate independent tuning of quantum dots in photonic crystal cavities.  We have shown that the cavities themselves are coupled through a waveguide by performing spatial scanning measurements. This method holds promise as an important component of a scalable photonic crystal-based quantum information architecture and for the implementation of quantum simulators based on coupled cavities.

The authors would like to acknowledge D. Kleckner for useful discussions.  This work was supported by NSF NIRT Grant No. 0304678 and Marie Curie EXT-CT-2006-042580.  A portion of this work was done in the UCSB nanofabrication facility, part of the NSF funded NNIN network.



\begin{thebibliography}{17}
\expandafter\ifx\csname natexlab\endcsname\relax\def\natexlab#1{#1}\fi
\expandafter\ifx\csname bibnamefont\endcsname\relax
  \def\bibnamefont#1{#1}\fi
\expandafter\ifx\csname bibfnamefont\endcsname\relax
  \def\bibfnamefont#1{#1}\fi
\expandafter\ifx\csname citenamefont\endcsname\relax
  \def\citenamefont#1{#1}\fi
\expandafter\ifx\csname url\endcsname\relax
  \def\url#1{\texttt{#1}}\fi
\expandafter\ifx\csname urlprefix\endcsname\relax\def\urlprefix{URL }\fi
\providecommand{\bibinfo}[2]{#2}
\providecommand{\eprint}[2][]{\url{#2}}

\bibitem[{\citenamefont{Childress et~al.}(2006)\citenamefont{Childress, Taylor,
  S\"{o}rensen, and Lukin}}]{ChildressPRL06}
\bibinfo{author}{\bibfnamefont{L.}~\bibnamefont{Childress}},
  \bibinfo{author}{\bibfnamefont{J.~M.} \bibnamefont{Taylor}},
  \bibinfo{author}{\bibfnamefont{A.~S.} \bibnamefont{S\"{o}rensen}},
  \bibnamefont{and} \bibinfo{author}{\bibfnamefont{M.~D.} \bibnamefont{Lukin}},
  \bibinfo{journal}{Phys. Rev. Lett.} \textbf{\bibinfo{volume}{96}},
  \bibinfo{pages}{070504} (\bibinfo{year}{2006}).

\bibitem[{\citenamefont{Cirac et~al.}({1997})\citenamefont{Cirac, Zoller,
  Kimble, and Mabuchi}}]{CiracPRL97}
\bibinfo{author}{\bibfnamefont{J.~I.} \bibnamefont{Cirac}},
  \bibinfo{author}{\bibfnamefont{P.}~\bibnamefont{Zoller}},
  \bibinfo{author}{\bibfnamefont{H.~J.} \bibnamefont{Kimble}},
  \bibnamefont{and} \bibinfo{author}{\bibfnamefont{H.}~\bibnamefont{Mabuchi}},
  \bibinfo{journal}{{Phys. Rev. Lett.}} \textbf{\bibinfo{volume}{{78}}},
  \bibinfo{pages}{{3221}} (\bibinfo{year}{{1997}}).

\bibitem[{\citenamefont{Bonato et~al.}(2010)\citenamefont{Bonato, Haupt,
  Oemrawsingh, Gudat, Ding, van Exter, and Bouwmeester}}]{bonatoPRL10}
\bibinfo{author}{\bibfnamefont{C.}~\bibnamefont{Bonato}},
  \bibinfo{author}{\bibfnamefont{F.}~\bibnamefont{Haupt}},
  \bibinfo{author}{\bibfnamefont{S.~S.~R.} \bibnamefont{Oemrawsingh}},
  \bibinfo{author}{\bibfnamefont{J.}~\bibnamefont{Gudat}},
  \bibinfo{author}{\bibfnamefont{D.}~\bibnamefont{Ding}},
  \bibinfo{author}{\bibfnamefont{M.~P.} \bibnamefont{van Exter}},
  \bibnamefont{and}
  \bibinfo{author}{\bibfnamefont{D.}~\bibnamefont{Bouwmeester}},
  \bibinfo{journal}{Phys. Rev. Lett.} \textbf{\bibinfo{volume}{104}},
  \bibinfo{pages}{160503} (\bibinfo{year}{2010}).

\bibitem[{\citenamefont{Cho and Lee}(2005)}]{choPRL05}
\bibinfo{author}{\bibfnamefont{J.}~\bibnamefont{Cho}} \bibnamefont{and}
  \bibinfo{author}{\bibfnamefont{H.-W.} \bibnamefont{Lee}},
  \bibinfo{journal}{Phys. Rev. Lett.} \textbf{\bibinfo{volume}{95}},
  \bibinfo{pages}{160501} (\bibinfo{year}{2005}).

\bibitem[{\citenamefont{Cho et~al.}(2008)\citenamefont{Cho, Angelakis, and
  Bose}}]{choPRA08}
\bibinfo{author}{\bibfnamefont{J.}~\bibnamefont{Cho}},
  \bibinfo{author}{\bibfnamefont{D.~G.} \bibnamefont{Angelakis}},
  \bibnamefont{and} \bibinfo{author}{\bibfnamefont{S.}~\bibnamefont{Bose}},
  \bibinfo{journal}{Phys. Rev. A} \textbf{\bibinfo{volume}{78}},
  \bibinfo{pages}{022323} (\bibinfo{year}{2008}).

\bibitem[{\citenamefont{Hu et~al.}(2008)\citenamefont{Hu, Young, O'Brien,
  Munro, and Rarity}}]{rarityPRB08}
\bibinfo{author}{\bibfnamefont{C.~Y.} \bibnamefont{Hu}},
  \bibinfo{author}{\bibfnamefont{A.}~\bibnamefont{Young}},
  \bibinfo{author}{\bibfnamefont{J.~L.} \bibnamefont{O'Brien}},
  \bibinfo{author}{\bibfnamefont{W.~J.} \bibnamefont{Munro}}, \bibnamefont{and}
  \bibinfo{author}{\bibfnamefont{J.~G.} \bibnamefont{Rarity}},
  \bibinfo{journal}{Phys. Rev. B} \textbf{\bibinfo{volume}{78}},
  \bibinfo{pages}{085307} (\bibinfo{year}{2008}).

\bibitem[{\citenamefont{Hartmann et~al.}(2006)\citenamefont{Hartmann, Brandao,
  and Plenio}}]{hartmannNP06}
\bibinfo{author}{\bibfnamefont{M.~J.} \bibnamefont{Hartmann}},
  \bibinfo{author}{\bibfnamefont{F.~G. S.~L.} \bibnamefont{Brandao}},
  \bibnamefont{and} \bibinfo{author}{\bibfnamefont{M.~B.}
  \bibnamefont{Plenio}}, \bibinfo{journal}{Nat Phys}
  \textbf{\bibinfo{volume}{2}}, \bibinfo{pages}{849} (\bibinfo{year}{2006}).

\bibitem[{\citenamefont{Hartmann et~al.}(2008)\citenamefont{Hartmann, Brandao,
  and Plenio}}]{hartmannLPR08}
\bibinfo{author}{\bibfnamefont{M.~J.} \bibnamefont{Hartmann}},
  \bibinfo{author}{\bibfnamefont{F.~G. S.~L.} \bibnamefont{Brandao}},
  \bibnamefont{and} \bibinfo{author}{\bibfnamefont{M.~B.}
  \bibnamefont{Plenio}}, \bibinfo{journal}{Laser \& Photon. Rev.}
  \textbf{\bibinfo{volume}{2}}, \bibinfo{pages}{527} (\bibinfo{year}{2008}).

\bibitem[{\citenamefont{Faraon et~al.}(2008)\citenamefont{Faraon, Fushman,
  Englund, Stoltz, Petroff, and Vu\v{c}kovi\'{c}}}]{FaraonOE08}
\bibinfo{author}{\bibfnamefont{A.}~\bibnamefont{Faraon}},
  \bibinfo{author}{\bibfnamefont{I.}~\bibnamefont{Fushman}},
  \bibinfo{author}{\bibfnamefont{D.}~\bibnamefont{Englund}},
  \bibinfo{author}{\bibfnamefont{N.}~\bibnamefont{Stoltz}},
  \bibinfo{author}{\bibfnamefont{P.}~\bibnamefont{Petroff}}, \bibnamefont{and}
  \bibinfo{author}{\bibfnamefont{J.}~\bibnamefont{Vu\v{c}kovi\'{c}}},
  \bibinfo{journal}{Opt. Express} \textbf{\bibinfo{volume}{16}},
  \bibinfo{pages}{12154} (\bibinfo{year}{2008}).

\bibitem[{\citenamefont{Faraon et~al.}(2007)\citenamefont{Faraon, Waks,
  Englund, Fushman, and Vu\v{c}kovi\'{c}}}]{FaraonAPL07b}
\bibinfo{author}{\bibfnamefont{A.}~\bibnamefont{Faraon}},
  \bibinfo{author}{\bibfnamefont{E.}~\bibnamefont{Waks}},
  \bibinfo{author}{\bibfnamefont{D.}~\bibnamefont{Englund}},
  \bibinfo{author}{\bibfnamefont{I.}~\bibnamefont{Fushman}}, \bibnamefont{and}
  \bibinfo{author}{\bibfnamefont{J.}~\bibnamefont{Vu\v{c}kovi\'{c}}},
  \bibinfo{journal}{Appl. Phys. Lett.} \textbf{\bibinfo{volume}{90}},
  \bibinfo{pages}{073102} (\bibinfo{year}{2007}).

\bibitem[{\citenamefont{Fushman et~al.}(2007)\citenamefont{Fushman, Waks,
  Englund, Stoltz, Petroff, and Vu\v{c}kovi\'{c}}}]{FushmanAPL07}
\bibinfo{author}{\bibfnamefont{I.}~\bibnamefont{Fushman}},
  \bibinfo{author}{\bibfnamefont{E.}~\bibnamefont{Waks}},
  \bibinfo{author}{\bibfnamefont{D.}~\bibnamefont{Englund}},
  \bibinfo{author}{\bibfnamefont{N.}~\bibnamefont{Stoltz}},
  \bibinfo{author}{\bibfnamefont{P.}~\bibnamefont{Petroff}}, \bibnamefont{and}
  \bibinfo{author}{\bibfnamefont{J.}~\bibnamefont{Vu\v{c}kovi\'{c}}},
  \bibinfo{journal}{Appl. Phys. Lett.} \textbf{\bibinfo{volume}{90}},
  \bibinfo{pages}{091118} (\bibinfo{year}{2007}).

\bibitem[{\citenamefont{Faraon et~al.}({2007})\citenamefont{Faraon, Englund,
  Fushman, Vu\v{c}kovi\'{c}, Stoltz, and Petroff}}]{FaraonAPL07}
\bibinfo{author}{\bibfnamefont{A.}~\bibnamefont{Faraon}},
  \bibinfo{author}{\bibfnamefont{D.}~\bibnamefont{Englund}},
  \bibinfo{author}{\bibfnamefont{I.}~\bibnamefont{Fushman}},
  \bibinfo{author}{\bibfnamefont{J.}~\bibnamefont{Vu\v{c}kovi\'{c}}},
  \bibinfo{author}{\bibfnamefont{N.}~\bibnamefont{Stoltz}}, \bibnamefont{and}
  \bibinfo{author}{\bibfnamefont{P.}~\bibnamefont{Petroff}},
  \bibinfo{journal}{{Appl. Phys. Lett.}} \textbf{\bibinfo{volume}{{90}}},
  \bibinfo{pages}{{213110}} (\bibinfo{year}{{2007}}).

\bibitem[{\citenamefont{Faraon et~al.}(2011)\citenamefont{Faraon, Majumdar,
  Englund, Kim, Bajcsy, and Vučković}}]{faraonNJP11}
\bibinfo{author}{\bibfnamefont{A.}~\bibnamefont{Faraon}},
  \bibinfo{author}{\bibfnamefont{A.}~\bibnamefont{Majumdar}},
  \bibinfo{author}{\bibfnamefont{D.}~\bibnamefont{Englund}},
  \bibinfo{author}{\bibfnamefont{E.}~\bibnamefont{Kim}},
  \bibinfo{author}{\bibfnamefont{M.}~\bibnamefont{Bajcsy}}, \bibnamefont{and}
  \bibinfo{author}{\bibfnamefont{J.}~\bibnamefont{Vučković}},
  \bibinfo{journal}{New Journal of Physics} \textbf{\bibinfo{volume}{13}},
  \bibinfo{pages}{055025} (\bibinfo{year}{2011}).

\bibitem[{\citenamefont{Ridgway et~al.}(1993)\citenamefont{Ridgway, Elliman,
  and Hauser}}]{ridgwayNIM93}
\bibinfo{author}{\bibfnamefont{M.}~\bibnamefont{Ridgway}},
  \bibinfo{author}{\bibfnamefont{R.}~\bibnamefont{Elliman}}, \bibnamefont{and}
  \bibinfo{author}{\bibfnamefont{N.}~\bibnamefont{Hauser}},
  \bibinfo{journal}{Nuclear Instruments and Methods in Physics Research Section
  B: Beam Interactions with Materials and Atoms}
  \textbf{\bibinfo{volume}{80-81}}, \bibinfo{pages}{835}
  (\bibinfo{year}{1993}).

\bibitem[{\citenamefont{Ziegler et~al.}(1985)\citenamefont{Ziegler, Biersack,
  and Littmark}}]{Ziegler85}
\bibinfo{author}{\bibfnamefont{J.~F.} \bibnamefont{Ziegler}},
  \bibinfo{author}{\bibfnamefont{J.~P.} \bibnamefont{Biersack}},
  \bibnamefont{and} \bibinfo{author}{\bibfnamefont{U.}~\bibnamefont{Littmark}},
  \emph{\bibinfo{title}{The Stopping and Range of Ions in Solids}}
  (\bibinfo{publisher}{Pergamon Press}, \bibinfo{address}{New York},
  \bibinfo{year}{1985}).

\bibitem[{\citenamefont{{Thon} et~al.}(2009)\citenamefont{{Thon}, {Rakher},
  {Kim}, {Gudat}, {Irvine}, {Petroff}, and {Bouwmeester}}}]{ThonAPL09}
\bibinfo{author}{\bibfnamefont{S.~M.} \bibnamefont{{Thon}}},
  \bibinfo{author}{\bibfnamefont{M.~T.} \bibnamefont{{Rakher}}},
  \bibinfo{author}{\bibfnamefont{H.}~\bibnamefont{{Kim}}},
  \bibinfo{author}{\bibfnamefont{J.}~\bibnamefont{{Gudat}}},
  \bibinfo{author}{\bibfnamefont{W.~T.~M.} \bibnamefont{{Irvine}}},
  \bibinfo{author}{\bibfnamefont{P.~M.} \bibnamefont{{Petroff}}},
  \bibnamefont{and}
  \bibinfo{author}{\bibfnamefont{D.}~\bibnamefont{{Bouwmeester}}},
  \bibinfo{journal}{Appl. Phys. Lett.} \textbf{\bibinfo{volume}{94}},
  \bibinfo{pages}{111115} (\bibinfo{year}{2009}).

\bibitem[{\citenamefont{Schneider et~al.}(2008)\citenamefont{Schneider,
  Strau\ss, S\"{u}nner, Huggenberger, Wiener, Reitzenstein, Kamp, H\"{o}fling,
  and Forchel}}]{schneiderAPL08}
\bibinfo{author}{\bibfnamefont{C.}~\bibnamefont{Schneider}},
  \bibinfo{author}{\bibfnamefont{M.}~\bibnamefont{Strau\ss}},
  \bibinfo{author}{\bibfnamefont{T.}~\bibnamefont{S\"{u}nner}},
  \bibinfo{author}{\bibfnamefont{A.}~\bibnamefont{Huggenberger}},
  \bibinfo{author}{\bibfnamefont{D.}~\bibnamefont{Wiener}},
  \bibinfo{author}{\bibfnamefont{S.}~\bibnamefont{Reitzenstein}},
  \bibinfo{author}{\bibfnamefont{M.}~\bibnamefont{Kamp}},
  \bibinfo{author}{\bibfnamefont{S.}~\bibnamefont{H\"{o}fling}},
  \bibnamefont{and} \bibinfo{author}{\bibfnamefont{A.}~\bibnamefont{Forchel}},
  \bibinfo{journal}{Applied Physics Letters} \textbf{\bibinfo{volume}{92}},
  \bibinfo{eid}{183101} (pages~\bibinfo{numpages}{3}) (\bibinfo{year}{2008}).

\end{thebibliography}
\end{document}